# AN EXTENSION OF SEMANTIC PROXIMITY FOR FUZZY MULTIVALUED DEPENDENCIES IN FUZZY RELATIONAL DATABASE

Arezoo Rajaei [1], Ahmad Baraani Dastjerdi [2] and Nasser Ghasem Aghaee [3]

[1] Department of Computer Engineering, Sheikhbahaee University, Isfahan, Iran
`arezoo_rajaee2000@yahoo.com`
[2] Department of Computer Engineering, Isfahan University, Isfahan, Iran
`ahmadb@eng.ui.ac.ir`
[3] Department of Computer Engineering, Sheikhbahaee University, Isfahan, Iran
`nasser_ga@yahoo.ca`

## ABSTRACT

*Following the development of fuzzy logic theory by Lotfi Zadeh, its applications were investigated by researchers in different fields. Presenting and working with uncertain data is a complex problem. To solve for such a complex problem, the structure of relationships and operators dependent on such relationships must be repaired. The fuzzy database has integrity limitations including data dependencies. In this paper, first fuzzy multivalued dependency based semantic proximity and its problems are studied. To solve these problems, the semantic proximity's formula is modified, and fuzzy multivalued dependency based on the concept of extension of semantic proximity with α degree is defined in fuzzy relational database which includes Crisp, NULL and fuzzy values, and also inference rules for this dependency are defined, and their completeness is proved. Finally, we will show that fuzzy functional dependency based on this concept is a special case of fuzzy multivalued dependency in fuzzy relational database.*

## 1. INTRODUCTION

Database technology is one of the most important and precedented technologies being used in the real world applications. Database is a computer system software tool for maintaining records, their updating, and retrieval [1].

There could be three problems involved in databases [1]:

- **Redundancy of data:** In relational databases, the frequency of data is the only way for connection between tables and be called foreign key. The frequency of data is irregular.

- **Abnormal**: The redundancy of data causes abnormality in the database.

- **NULL values:** Null values occupy many spaces in databases.

Because of these problems, there is a normalization process for normalizing the database.

The goal of normalizing is to eliminate data redundancy as well as to maintain dependency between the corresponding data. This is to reduce the size of the database and ensure logical storing of the data [1, 13].

             157

Normalization process will prevent anomalies due to updating changes in the database. Applying normalization process will result in an efficient and reliable database. Following the normalization concept, data dependency concept was developed.

Classical multivalued dependency is one of data dependencies in classical relational databases that is used for normalizing operation in these databases. Multivalued dependency (MVD) means that the presence of certain records in a table implies the existence of other certain records [1, 17].

MVD is more general than FD such that each FD is a MVD.

In 1965 at the University of California, Berkeley, Lotfi Zadeh introduced the theory of fuzzy sets and fuzzy logic, two concepts that laid the foundation for possibility theory in 1977 [2]. One of these fields is related to the fuzzy theory application in database systems, information retrieval, and expert system and knowledge base.

In fact, the fuzzy relational data model is an extension of classical relational data that records ambiguous data values and their dependencies. Also the most modern computer systems are based on this model. The relational modal was first proposed by Codd [3, 12]. The basic model of fuzzy relational databases is considered to be the simplest one and it consists of adding a grade, normally in the [0, 1] interval, to each instance (or tuple). This makes keeping the database data homogeneity possible [2].

In the different fuzzy relational database models, tasks are accomplished for expressing data dependencies specially FMVD and FFD [4, 6, 14, 15, 16, 18].

One of the fuzzy concepts in fuzzy relational data model is a semantic proximity concept and based on this concept, the fuzzy values are defined as the interval numbers. Also fuzzy data dependencies including FFDs and FMVDs have been defined based on this concept [4].

Since in fuzzy database Crisp and NULL values are existed in addition to fuzzy values, a method exists for converting non-fuzzy values to interval numbers that eliminate non-fuzzy values from this database.

FMVD based on semantic proximity concept has problems [5]. In this paper we discuss how to improve semantic proximity formula. This new formula will solve all FMVD problems.

It is important to note that fuzzy values in the form of interval numbers might still have some out of the range data. However, fuzzy sets at degree α [11], that cover all possible ranges of fuzzy values, may flexibly give ranges to fuzzy values that support the needs of database designers. Therefore, the semantic proximity concept is extended [6].

FMVD based on the extension of semantic proximity at α degree is defined for fuzzy relational databases including Fuzzy, Crisp and NULL values. Also inference rules are defined for FMVD and FFD-FMVD and it will be shown FFD and MVD to be special cases of FMVD.

In the second section of this paper we will summarize previous works in fuzzy relational databases. In the third section we will discuss how Crisp and NULL values can be transformed to interval numbers. In the fourth section we present a new definition for the semantic proximity concept for solving the FMVD's problem. In the fifth section we define FMVD and its inference rules based on the extension of semantic proximity concept. Finally we will prove this inference rules to be complete and will provide some certain states from this dependency.

## 2. PREVIOUS WORK

Semantic proximity proposed by Liu [4] with a form of interval numbers, utilizing semantic relation. Also Liu [4] defined FFD, FMVD, and FJD with their inference rules, but he didn't explain how Crisp and NULL values can transform to interval numbers for FMVD.

Danga and Tran [5] discussed definitions of FFD and FMVD given by Liu [4], and defined a new definition for FMVD. But this definition doesn't work for some cases, which have Crisp and NULL values in fuzzy databases.

Liao, Wang and Liu [10] described Crisp and NULL values and defined FD based on this description and the semantic proximity concept.

Furthermore, Lee and Pang [6] extended semantic proximity with the concept of fuzzy sets at a degree α set by database designers, and determined FFDs. They proved the inference rules to be sound and complete, but used only fuzzy values based on interval number. They didn't define FMVD and didn't conclude that if a fuzzy database satisfies FFD condition then it satisfies FMVD condition as well.

## 3. TRANSFORMING CRISP AND NULL VALUES TO INTERVAL NUMBERS

A fuzzy subset X in dom(Ai) is characterized by an ordered couple [a, b]/p. [a, b] is called the interval number, where a and b are real numbers. P ($0 \leq p \leq 1$) is the degree of confidence [4].

The confidence degree of a crisp value is one. For example, 3.6 is denoted [3.6, 3.6]. The domain of a NULL value is the entire universe of discourse. The NULL value is denoted [L, U] where U is the upper bound and L is the lower bound [10].

## 4. MODIFYING THE FORMULA OF SEMANTIC PROXIMITY

As defined [4], the semantic proximity between two fuzzy values $f_1$ and $f_2$ can be described by the following definitions:

***Definition 1(Semantic Proximity)*:** If there are two fuzzy values $f_1$ and $f_2$, then $SP(f_1,f_2)$ is defined as:

$$SP(f_1,f_2) = \frac{\|f_1 \cap f_2\|}{\|f_1 \cup f_2\|} - \frac{\|f_1 \cap f_2\|}{\alpha}, \quad (0 \leq SP(f_1,f_2) \leq 1) \quad (1)$$

And $\|h\|$ is the modular of interval h.

$$\|h\| = \begin{cases} 0 & h = 0, \\ \delta & h = [a,a], \\ |b-a| & h = [a,b] \text{ and } a \neq b, \\ \alpha & h = \infty. \end{cases}$$

where a and b are real numbers, and $\alpha$ is a given coefficient about the universe of the discourse, $\alpha \geq \|f_1 \cup f_2\|$. δ is relatively very small. So we select α/10000 for δ [4].

The following properties ought to be satisfied by SP ($f_1$, $f_2$):

Let $f_1 = [a_1, b_1]$, $f_2 = [a_2, b_2]$, $g_1 = [c_1, d_1]$, $g_2 = [c_2, d_2]$.

1. SP ($f_1$, $f_2$) =1 if and only if $a_1 = b_1 = a_2 = b_2$.

2. SP ($f_1$, $f_2$) =0 if and only if $f_1 \cap f_2 = \phi$.

3. If $a_1=a_2$, $b_1=b_2$, $c_1=c_2$, $d_1=d_2$, and $|d_1-c_1|=|b_1-a_1|$ then SP ($f_1$, $f_2$)>=SP ($g_1$, $g_2$).

4. If $|a_2-b_2|=|a_1-b_1|$ and $f_1 \cap g_1 \geq f_1 \cap g_1$ then SP ($f_1$, $g_1$)>=SP ($f_2$, $g_1$).

The definition of FFD and FMVD based on the semantic proximity concept are presented in [4, 5]:

***Definition 2(Fuzzy Functional Dependency)*:** A FFD X~-> Y with X; Y $\subset$ U holds in a fuzzy instance r on U, if for all $T_i$ and $T_j \in r$ we have $SP(T_i[X],T_j[X]) \leq SP(T_i[Y],T_j[Y])$ [4,5].

***Definition 3(Fuzzy Multivalued Dependency):*** Let X; Y $\subset$ U and Z =U − XY. A FMVD X$\sim\to\sim\to$Y holds in a fuzzy instance r on U if, for any two tuples $T_i$ ; $T_j \in$ r with $SP(T_i[X],T_j[X])=\alpha$, there exists a tuple T in r with $SP(T[X],T_i[X])\geq\alpha$, $SP(T[X],T_j[X])\geq\alpha$, $SP(T[Y],T_i[Y])\geq\alpha$ and $SP(T[Z],T_j[Z])\geq\alpha$ [5].

However, the formula (1) sometimes doesn't work in fuzzy relational database which have Crisp and NULL values. We shall discuss the following example in Table 1. This table lists Crisp and fuzzy values where X, Y represent Crisp values, and Z stands for fuzzy value.

Table 1: a table with Crisp and fuzzy values

|    | X      | Y      | Z        |
|----|--------|--------|----------|
| T1 | [a, a] | [b, b] | [c1, d1] |
| T2 | [a, a] | [b, b] | [c2, d2] |

Consider a relation $R_1$. $R_1$ satisfies X$\sim$--> Y because $SP(T_i[X],T_j[X]) \leq SP(T_i[Y],T_j[Y])$, but it doesn't satisfy FMVD. Consider two tuples $t_1$ and $t_2$, we have $SP(T_i[X],T_j[X])=1$. For all $t_3 \in R_1$ (not necessary distinct from $t_1$ and $t_2$), $SP(T[X],T_i[X])=1$, $SP(T[X],T_j[X])=1$, $SP(T[Y],T_i[Y])=1$, and $SP(T[Z],T_j[Z]) \leq 1$ because $SP([1,9],[1,9]) \prec 1$, $SP([1,8],[1,8]) \prec 1$, and $SP([1,9],[1,8]) \prec 1$.

Based on the replication rule, if X $\sim$--> Y, then X$\sim$-->$\sim$--> Y, R must satisfy FMVD. So we must fix the formula for solving this problem.

There are many expressions of the semantic proximity in in [7, 8]. We hereby present one of them as described below.

***Definition 4(semantic proximity):*** If $f_1$= [$a_1$, $b_1$], $f_2$= [$a_2$, $b_2$] are two fuzzy values, and $a_1$, $a_2$, $b_1$ and $b_2$ are real numbers, then the semantic proximity between $f_1$ and $f_2$ is shown as formula 2:

$$SP(f_1,f_2) = 1 - (p(a_1,a_2) + p(b_1,b_2)) \quad (2)$$

where,

$$p(a_1,a_2) = \frac{|a_1 - a_2|}{MAX(a_1,a_2)}$$

As is considered, the formula 2 obtains from the complement of semantic distance (SD) between $f_1$ and $f_2$.

$$SP(f_1,f_2) = 1 - SD(f_1,f_2) \quad (3)$$

The above relation (2) doesn't have to hold for properties (2) and (4) of semantic proximity. This is due to the fact that if $f_1 \cap f_2 = \phi$, then $SP(f_1,f_2)$ doesn't equal to zero. Therefore, we improve on the relation (2) of semantic proximity for solving this problem.

***Definition 5(improved semantic proximity):*** if $f_1$= [$a_1$, $b_1$] and $f_2$= [$a_2$, $b_2$] are two fuzzy values, and $a_1$, $a_2$, $b_1$ and $b_2$ are real numbers, then the semantic proximity between $f_1$ and $f_2$ is shown as (4):

$$\text{If } (f_1 \cap f_2) = \phi \text{ then } SP(f_1,f_2) = 0$$
$$\text{Else } SP(f_1,f_2) = 1 - (p(a_1,a_2) + p(b_1,b_2)) \quad (4)$$

where,

$$p(a_1, a_2) = \frac{|a_1 - a_2|}{MAX(a_1, a_2)}$$

This definition solves the problems in fuzzy relational databases which have Crisp, NULL, and fuzzy values. It is also valid for all of semantic proximity properties.

## 5. DEFINING FMVD BASED ON THE EXTENSION OF SEMANTIC PROXIMITY

Lee and Pang [6] defined the extension of semantic proximity as following:

***Definition 6(the extension of semantic proximity):*** Let $\tilde{G}$ and $\tilde{H}$ be fuzzy sets. The semantic proximity between $\tilde{G}^\alpha$ and $\tilde{H}^\alpha$ at degree α is defined as:

$$SP^\alpha(\tilde{G}^\alpha, \tilde{H}^\alpha) = \frac{d(\tilde{G}^\alpha \cap \tilde{H}^\alpha)}{d(\tilde{G}^\alpha \cup \tilde{H}^\alpha)} - \frac{d(\tilde{G}^\alpha \cap \tilde{H}^\alpha)}{\theta} \quad (5)$$

where for $\alpha \in [0, 1]$ and for closed interval $[x, y]$, $d([x, y])$ is defined as:

$$d([x,y]) = \begin{cases} 0 & \text{if } d([x,y]) = \phi, \\ \varepsilon & \text{if } x = y, \\ |y - x| & \text{if } x \neq y, \\ \alpha & \text{otherwise} \end{cases}$$

The degree α, set by database designers, determines the ranges of fuzzy sets. The smaller the range, the harder to determine the strength of semantic proximity. θ is the scope of the universe, $\theta \geq [\tilde{G}^\alpha, \tilde{H}^\beta]$, ε is a relatively small positive number, and x,y are real numbers[6].

Let $\tilde{G}^\alpha = [w_i, x_i]$ and $\tilde{H}^\alpha = [y_i, z_i]$ (for i = 1, 2) be the fuzzy sets at degree α for $\alpha \in [0,1]$. The length of $\tilde{G}_i^\alpha$ and $\tilde{H}_i^\alpha$ are $|x_i - w_i|$ and $|z_i - y_i|$, respectively.

The properties of semantic proximity are [6]:

1. If $w_i = x_i$ for i = 1, 2, which means $\tilde{G}_i^\alpha$ and $\tilde{H}_i^\alpha$ become crisp data, then $SP^\alpha(\tilde{G}_1^\alpha, \tilde{G}_2^\alpha) = 1$.

2. If $\tilde{G}^\alpha \cap \tilde{H}^\alpha = \phi$, then $SP^\alpha(\tilde{G}^\alpha, \tilde{H}^\alpha) = 0$.

3. If $\tilde{G}_1^\alpha = \tilde{G}_2^\alpha$, $\tilde{H}_1^\alpha = \tilde{H}_2^\alpha$, and $|z_1 - y_1| \geq |x_1 - w_1|$, then $SP^\alpha(\tilde{G}_1^\alpha, \tilde{G}_2^\alpha) \geq SP^\alpha(\tilde{H}_1^\alpha, \tilde{H}_2^\alpha)$.

4. If $\tilde{G}_1^\alpha = \tilde{G}_2^\alpha$, and $\tilde{G}_1^\alpha \cap \tilde{H}^\alpha \geq \tilde{G}_2^\alpha \cap \tilde{H}^\alpha$, then $SP^\alpha(\tilde{G}_1^\alpha, \tilde{H}^\alpha) \geq SP^\alpha(\tilde{G}_2^\alpha, \tilde{H}^\alpha)$.

The definition may be generalized to measure the semantic proximity of two tuples in fuzzy relational databases. Let $\tilde{A}_{ij}$ be fuzzy sets for i = 1, 2, j =1, 2,..., n, and $t_1 = (\tilde{A}_{11}, \tilde{A}_{12},..., \tilde{A}_{1n})$ and $t_2 = (\tilde{A}_{21}, \tilde{A}_{22},..., \tilde{A}_{2n})$ form two tuples in a fuzzy relational instance r. The semantic proximity of two tuples at degree α is:

$$SP^\alpha(t_1^\alpha, t_2^\alpha) = \min_{1 \leq j \leq n} SP^\alpha(\tilde{A}_{1j}^\alpha, \tilde{A}_{2j}^\alpha), \; For \; \alpha \in [0,1] \, [6].$$

Lee and Pang [6] defined FFD at degree α with its inference rules. We know in every relational database if each table satisfies FD condition, then it satisfies MVD condition. Now we like to investigate if each table satisfies FFD condition at degree α, would it also satisfy FMVD at degree α. Therefore, we must define FMVD at degree *α* based on the extension of semantic proximity and the inspiration of definition 3.

161

***Definition 7(FMVD based on the extension of semantic proximity)***: Let U be a relation scheme, let X, Y⊆U, and Z = U - (X Y). A relation r(U) satisfies the fuzzy multivalued dependency (FMVD), $X \sim\to\sim\to_\alpha Y$, if for any two tuples $T_i$ and $T_j$ in r with $SP^\alpha(T_i^\alpha[X], T_j^\alpha[X]) = \beta$, there exists a tuple T in r with $SP^\alpha(T^\alpha[X], T_i^\alpha[X]) \geq \beta$, $SP^\alpha(T^\alpha[X], T_j^\alpha[X]) \geq \beta$, $SP^\alpha(T^\alpha[Y], T_i^\alpha[Y]) \geq \beta$, and $SP^\alpha(T^\alpha[Z], T_j^\alpha[Z]) \geq \beta$.

The main difference between definitions 7 and 3 is the applied degree α because fuzzy sets at degree α cover all possible ranges of fuzzy values. Therefore, working with these fuzzy sets will be much easier.

***Lemma 1***: If relation r on scheme R satisfies the FMVD $X \sim\to\sim\to_\alpha Y$ and Z=R - (X Y), then r satisfies $X \sim\to\sim\to_\alpha Z$.

***Proof***: Based on a lemma for MVD [10], If relation r on scheme R satisfies the MVD $X \to\to Y$ and Z=R - (X Y), then r satisfies $X \to\to Z$. We can prove lemma 1 as below.

We consider two cases: The first case, suppose X and Y are not disjoint, r satisfies $X \to\to Y$, $Y' = Y - X$ and $Z = U - XY = U - XY'$. According to the above lemma, r must satisfy $X \sim\to\sim\to_\alpha Y'$. If two tuples $T_1$ and $T_2$ in r with $SP^\alpha(T_1^\alpha[X], T_2^\alpha[X]) = \beta$, and $X \sim\to\sim\to_\alpha Y$, there must exist a tuple T in r with $SP^\alpha(T^\alpha[X], T_1^\alpha[X]) \geq \beta$, $SP^\alpha(T^\alpha[X], T_2^\alpha[X]) \geq \beta$, $SP^\alpha(T^\alpha[Y], T_1^\alpha[Y]) \geq \beta$, and $SP^\alpha(T^\alpha[Z], T_2^\alpha[Z]) \geq \beta$. Due to the fact that $Y' \subseteq Y$, if $SP^\alpha(T^\alpha[Y], T_1^\alpha[Y]) \geq \beta$, then $SP^\alpha(T^\alpha[Y'], T_1^\alpha[Y']) \geq \beta$. So r will satisfy $X \sim\to\sim\to_\alpha Y'$.

The second case, suppose X and Y are disjoint, r satisfies $X \to\to Y$, and $X' \subseteq X$. According to the above lemma, r must satisfy $X \sim\to\sim\to_\alpha YX'$. If two tuples $T_1$ and $T_2$ in r with $SP^\alpha(T_1^\alpha[X], T_2^\alpha[X]) = \beta$, and $X \sim\to\sim\to_\alpha Y$, there must exist a tuple T in r with $SP^\alpha(T^\alpha[X], T_1^\alpha[X]) \geq \beta$, $SP^\alpha(T^\alpha[X], T_2^\alpha[X]) \geq \beta$, $SP^\alpha(T^\alpha[Y], T_1^\alpha[Y]) \geq \beta$, and $SP^\alpha(T^\alpha[Z], T_2^\alpha[Z]) \geq \beta$. Since $X' \subseteq X$, if $SP^\alpha(T^\alpha[X], T_1^\alpha[X]) \geq \beta$, then $SP^\alpha(T^\alpha[X'], T_1^\alpha[X']) \geq \beta$. Therefore r will satisfy $X \sim\to\sim\to_\alpha Y'$ because X and Y are disjoint.

***Theorem 1***: Let t be a relation on scheme R, and let X, Y, and Z be subsets of R such that Z=R-(X Y). Relation r satisfies the FMVD $X \sim\to\sim\to_\alpha Y$, if and only if r decomposes losslessly onto the relation schemes R1=XY and R2=XZ.

***Proof***: Based on a Theorem for MVD [10], relation r on scheme R satisfies the MVD $X \to\to Y$, if the only if r decomposes losslessly onto the relation schemes R1=XY and R2=XZ. We can prove the theorem 1 according to the following.

Let's assume the FMVD holds. Let $r_1 = \pi_{R_1}(r)$, and $r_2 = \pi_{R_2}(r)$. Let T be a tuple in $r_1 \infty r_2$. There must be a tuple $T_1 \in r_1$ and a tuple $T_2 \in r_2$ such that $SP^\alpha(T^\alpha[X], T_1^\alpha[X]) \geq \beta$, $SP^\alpha(T^\alpha[X], T_2^\alpha[X]) \geq \beta$, $SP^\alpha(T^\alpha[Y], T_1^\alpha[Y]) \geq \beta$ and

$SP^\alpha(T^\alpha[Z], T_2^\alpha[Z]) \geq \beta$. Since $r_1$ and $r_2$ are projections of r, there must be tuples $T'_1$ and $T'_2$ in T

with $SP^{\alpha}(T_1^{\alpha}[XY], T_1'^{\alpha}[XY]) \geq \beta$, and $SP^{\alpha}(T_2^{\alpha}[XZ], T_2'^{\alpha}[XZ]) \geq \beta$. The FMVD $X \sim\rightarrow\sim\rightarrow_{\alpha} Y$ implies that T must be in r, since r must contain a tuple $T_3$ with $SP^{\alpha}(T_3^{\alpha}[X], T_1'^{\alpha}[X]) \geq \beta$, $SP^{\alpha}(T_3^{\alpha}[X], T_2^{\alpha}[X]) \geq \beta$, $SP^{\alpha}(T_3^{\alpha}[Y], T_1'^{\alpha}[Y]) \geq \beta$ and $SP^{\alpha}(T_3^{\alpha}[Z], T_1'^{\alpha}[Z]) \geq \beta$, which is a description of.

Now suppose that r decomposes losslessly onto R1 and R2. Let $T_1$ and $T_2$ be tuples in r such that $SP^{\alpha}(T_1^{\alpha}[X], T_2^{\alpha}[X]) \geq \beta$. Let $r_1$ and $r_2$ be defined as before. Relation $r_1$ contains a tuple $T'_1 = T_1[XY]$ and relation $r_2$ contains a tuple $T'_2 = T_2[XZ]$. Since $r = r_1 \infty r_2$, r contains a tuple T such that $t SP^{\alpha}(T^{\alpha}[XY], T_1^{\alpha}[XY]) \geq \beta$ and $SP^{\alpha}(T^{\alpha}[XZ], T_2^{\alpha}[XZ]) \geq \beta$. Tuple T is the result of joining $T'_1$ and $T'_2$. Hence $T_1$ and $T_2$ cannot be used in a counter example, hence r satisfies $X \sim\rightarrow\sim\rightarrow_{\alpha} Y$.

From Theorem1 we can derive the following corollary.

*Corollary 1*: Let r be a relation on scheme R and let X and Y be subsets of R. If r satisfies the FFD $X \sim\rightarrow_{\alpha} Y$, then r satisfies the FMVD $X \sim\rightarrow\sim\rightarrow_{\alpha} Y$.

*Proof*: We know that $X \sim\rightarrow_{\alpha} Y$ implies that r decomposes lossless onto XY and X(R - (XY)). This result $X \sim\rightarrow\sim\rightarrow_{\alpha} Y$ is obtained directly from Theorem1.

*Theorem 2*: For all X, Y and U, X, Y⊆U, and for each relation is member of R(U), FMVD $X \sim\rightarrow\sim\rightarrow_{\alpha} Y$ is valid in R if and only if $X \sim\rightarrow\sim\rightarrow_{\alpha} Y - X$ is valid in R.

*Proof*: If X and Y are disjoint (X∩Y=ϕ), then $X \sim\rightarrow\sim\rightarrow_{\alpha} \phi$ always is valid, else if X and Y are not disjoint, then $Y - X \subseteq Y$ and since $X \sim\rightarrow\sim\rightarrow_{\alpha} Y$, then $X \sim\rightarrow\sim\rightarrow_{\alpha} Y - X$.

## 5.1. Inference rules for FMVD

FMVD inference rules based on extension of semantic proximity are inspired by FMVD inference rules [5] and MVD inference rules [10]. Suppose r is a relation on scheme R and W, X, Y, and Z are subsets of R.

1. *Complementation rule:* If r satisfies the FMVD $X \sim\rightarrow\sim\rightarrow_{\alpha} Y$ and $Z = R - (X Y)$, then r satisfies $X \sim\rightarrow\sim\rightarrow_{\alpha} Z$.

    *Proof*: According to Lemma 1, this rule is proved.

2. *Reflexivity rule*: If $Y \subseteq X$, then r satisfies $X \sim\rightarrow\sim\rightarrow_{\alpha} Y$.

    *Proof*: This rule is an immediate result of the definition of FMVD.

3. *Augmentation rule*: If r satisfies $X \sim\rightarrow\sim\rightarrow_{\alpha} Y$ and V⊆W, then r satisfies $XW \sim\rightarrow\sim\rightarrow_{\alpha} YV$.

    *Proof*: We know V⊆W. Therefore according to reflexivity rule, r should satisfy $W \sim\rightarrow\sim\rightarrow_{\alpha} V$. For any two tuples $T_i$ and $T_j$ in r with $SP^{\alpha}(T_i^{\alpha}[W], T_j^{\alpha}[W]) = \beta_1$, there exists a tuple T in r with the following conditions:

$SP^\alpha(T^\alpha[W], T_i^\alpha[W]) \geq \beta_1$
$SP^\alpha(T^\alpha[W], T_j^\alpha[W]) \geq \beta_1$
$SP^\alpha(T^\alpha[V], T_i^\alpha[V]) \geq \beta_1$ \qquad (I)
$SP^\alpha(T^\alpha[U - WV], T_j^\alpha[U - WV]) \geq \beta_1$

Also r satisfies $X \sim\to\sim\to_\alpha Y$, then for any two tuples $T_i$ and $T_j$ in r $SP^\alpha(T_i^\alpha[X], T_j^\alpha[X]) = \beta_2$, there exists a tuple T in r with the following conditions:

$SP^\alpha(T^\alpha[X], T_i^\alpha[X]) \geq \beta_2$
$SP^\alpha(T^\alpha[X], T_j^\alpha[X]) \geq \beta_2$
$SP^\alpha(T^\alpha[Y], T_i^\alpha[Y]) \geq \beta_2$ \qquad (II)
$SP^\alpha(T^\alpha[U - XY], T_j^\alpha[U - XY]) \geq \beta_2$

Upon merging (I) and (II), the following results are obtained:

$\min(SP^\alpha(T^\alpha[X], T_i^\alpha[X]), SP^\alpha(T^\alpha[W], T_i^\alpha[W])) \geq \max(\beta_1, \beta_2)$
$\min(SP^\alpha(T^\alpha[X], T_j^\alpha[X]), SP^\alpha(T^\alpha[W], T_j^\alpha[W])) \geq \max(\beta_1, \beta_2)$
$\min(SP^\alpha(T^\alpha[Y], T_i^\alpha[Y]), SP^\alpha(T^\alpha[V], T_i^\alpha[V])) \geq \max(\beta_1, \beta_2)$
$\min(SP^\alpha(T^\alpha[U - XY], T_j^\alpha[U - XY]), SP^\alpha(T^\alpha[U - WV], T_j^\alpha[U - WV])) \geq \max(\beta_1, \beta_2)$

The final results follow, assuming XW and YV are disjoint):

$SP^\alpha(T^\alpha[XW], T_i^\alpha[XW]) \geq \max(\beta_1, \beta_2)$
$SP^\alpha(T^\alpha[XW], T_j^\alpha[XW]) \geq \max(\beta_1, \beta_2)$
$SP^\alpha(T^\alpha[YV], T_i^\alpha[YV]) \geq \max(\beta_1, \beta_2)$ \qquad (III)
$SP^\alpha(T^\alpha[U - XWYV], T_j^\alpha[U - XWYV]) \geq \max(\beta_1, \beta_2)$

By substituting $\beta = \max(\beta_1, \beta_2)$ in (III), r will satisfy $XW \sim\to\sim\to_\alpha YV$.

4. *Additivity rule*: If r satisfies $X \sim\to\sim\to_\alpha Y$ and $X \sim\to\sim\to_\alpha Z$, then r satisfies $X \sim\to\sim\to_\alpha YZ$.

*Proof*: We know r satisfies $X \sim\to\sim\to_\alpha Y$. Therefore, for any two tuples $T_i$ and $T_j$ in r $SP^\alpha(T_i^\alpha[X], T_j^\alpha[X]) = \beta$, there exists a tuple T in r such that: ($Z = U - XY$)

$SP^\alpha(T^\alpha[X], T_i^\alpha[X]) \geq \beta$
$SP^\alpha(T^\alpha[X], T_j^\alpha[X]) \geq \beta$
$SP^\alpha(T^\alpha[Y], T_i^\alpha[Y]) \geq \beta$ \qquad (I)
$SP^\alpha(T^\alpha[Z], T_j^\alpha[Z]) \geq \beta$

Also r satisfies $X \sim\to\sim\to_\alpha Z$, then for any two tuples $T_i$ and $T_j$ in r $SP^\alpha(T_i^\alpha[X], T_j^\alpha[X]) = \beta$, there exists a tuple T in r such that:

164

$$SP^\alpha(T^\alpha[X], T_i^\alpha[X]) \geq \beta$$
$$SP^\alpha(T^\alpha[X], T_j^\alpha[X]) \geq \beta$$
$$SP^\alpha(T^\alpha[Z], T_i^\alpha[Z]) \geq \beta \quad \text{(II)}$$
$$SP^\alpha(T^\alpha[Y], T_j^\alpha[Y]) \geq \beta$$

Upon merging (I) and (II), the following statements can be made:

$$SP^\alpha(T^\alpha[X], T_i^\alpha[X]) \geq \beta$$
$$SP^\alpha(T^\alpha[X], T_j^\alpha[X]) \geq \beta$$
$$\min(SP^\alpha(T^\alpha[Y], T_i^\alpha[Y]), SP^\alpha(T^\alpha[Z], T_i^\alpha[Z])) \geq \beta \Rightarrow SP^\alpha(T^\alpha[YZ], T_i^\alpha[YZ]) \geq \beta$$
$$\min(SP^\alpha(T^\alpha[Y], T_j^\alpha[Y]), SP^\alpha(T^\alpha[Z], T_j^\alpha[Z])) \geq \beta \Rightarrow SP^\alpha(T^\alpha[YZ], T_j^\alpha[YZ]) \geq \beta$$

Then r satisfies $X \sim\rightarrow\sim\rightarrow_\alpha YZ$.

5. **Transitivity rule**: If r satisfies $X \sim\rightarrow\sim\rightarrow_\alpha Y$ and $Y \sim\rightarrow\sim\rightarrow_\alpha Z$, then r will satisfy $X \sim\rightarrow\sim\rightarrow_\alpha (Z-Y)$.

**Proof**: For proving this rule, first we must show if r satisfies $X \sim\rightarrow\sim\rightarrow_\alpha Y$ and $Y \sim\rightarrow\sim\rightarrow_\alpha Z$, then r will satisfy $X \sim\rightarrow\sim\rightarrow_\alpha YZ$. Suppose $W = R-(XYZ)$. We must show if any two tuples $T_i$ and $T_j$ with $SP^\alpha(T_1^\alpha[X], T_2^\alpha[X]) = \beta$ are in r, then, there exists a tuple T in r with these conditions:

$$SP^\alpha(T^\alpha[X], T_1^\alpha[X]) \geq \beta$$
$$SP^\alpha(T^\alpha[X], T_2^\alpha[X]) \geq \beta$$
$$SP^\alpha(T^\alpha[YZ], T_1^\alpha[YZ]) \geq \beta$$
$$SP^\alpha(T^\alpha[W], T_2^\alpha[W]) \geq \beta$$

(I)

From $X \sim\rightarrow\sim\rightarrow_\alpha Y$ we consider a tuple $T_3$ with its conditions: ($V = R-(XY)$)

$$SP^\alpha(T_3^\alpha[X], T_1^\alpha[X]) \geq \beta$$
$$SP^\alpha(T_3^\alpha[X], T_2^\alpha[X]) \geq \beta$$
$$SP^\alpha(T_3^\alpha[Y], T_1^\alpha[Y]) \geq \beta \quad \text{(II)}$$
$$SP^\alpha(T_3^\alpha[V], T_2^\alpha[V]) \geq \beta$$

From $Y \sim\rightarrow\sim\rightarrow_\alpha Z$ we consider a tuple $T_4$ with its conditions: ($U = R-(YZ)$)

$$SP^\alpha(T_4^\alpha[Y], T_1^\alpha[Y]) \geq \beta$$
$$SP^\alpha(T_4^\alpha[Y], T_2^\alpha[Y]) \geq \beta$$
$$SP^\alpha(T_4^\alpha[Z], T_1^\alpha[Z]) \geq \beta \quad \text{(III)}$$
$$SP^\alpha(T_4^\alpha[U], T_2^\alpha[U]) \geq \beta$$

We know $SP^\alpha(T_4^\alpha[X], T_1^\alpha[X]) \geq \beta$, since there is only one possible value for each attribute $A \in X$. Clearly $SP^\alpha(T_4^\alpha[YZ], T_1^\alpha[YZ]) \geq \beta$. Since $W \subseteq U - X \subseteq V$, therefore, we have $SP^{\alpha}(T_4^{\alpha}[W], T_2^{\alpha}[W]) \geq \beta$. Hence, $T_4$ is the tuple t we are seeking for. We have shown r satisfies $X \sim\to\sim\to_\alpha YZ$. Using Projectivity rule and $X \sim\to\sim\to_\alpha Y$, we finally get $X \sim\to\sim\to_\alpha (Z-Y)$.

6. **_Pseudo-transitivity rule_**: If r satisfies $X \sim\to\sim\to_\alpha Y$ and $YW \sim\to\sim\to_\alpha Z$, then r will satisfy $XW \sim\to\sim\to_\alpha (Z-YW)$.

    **_Proof_**: Using Additivity rule on $X \sim\to\sim\to_\alpha Y$, we get $XW \sim\to\sim\to_\alpha YW$. Therefore, by using Transitivity rule we finally get $XW \sim\to\sim\to_\alpha Z-YW$.

7. **_Projectivity rule_**: If r satisfies $X \sim\to\sim\to_\alpha Y$ and $X \sim\to\sim\to_\alpha Z$, then r will satisfy $X \sim\to\sim\to_\alpha (Y \cap Z)$, $X \sim\to\sim\to_\alpha (Y-Z)$, and $X \sim\to\sim\to_\alpha (Z-Y)$.

    **_Proof_**: By using Additivity rule, we can get $X \sim\to\sim\to_\alpha YZ$. Also $Y \cap Z \subseteq YZ$, $Y-Z \subseteq YZ$ and $Z-Y \subseteq YZ$ are in r, therefore, by using Reflexivity rule we obtain $YZ \sim\to\sim\to_\alpha (Y \cap Z)$, $YZ \sim\to\sim\to_\alpha (Y-Z)$ and $YZ \sim\to\sim\to_\alpha (Z-Y)$. By using Transitivity rule we finally get $X \sim\to\sim\to_\alpha (Y \cap Z)$, $X \sim\to\sim\to_\alpha (Y-Z)$, and $X \sim\to\sim\to_\alpha (Z-Y)$.

## 5.2. Inference rules for FMVD-FFD

There are only two rules for FDs and MVDs combined. Let r be a relation on R and let X, Y be subsets of R.

1. **_Replication rule_**: If r satisfies the FFD $X \sim\to_\alpha Y$, then r will satisfy the FMVD $X \sim\to\sim\to_\alpha Y$.

    **_Proof_**: This is a consequence of the corollary 1.

2. **_Coalescence rule_**: If r satisfies $X \sim\to\sim\to_\alpha Y$, $Z \sim\to_\alpha W$ where $W \subseteq Y$ and $Z \cap Y = \phi$, then r will satisfy $X \sim\to_\alpha W$.

    **_Proof_**: Let $T_1$ and $T_2$ be tuples in Y with $SP^{\alpha}(T_1^{\alpha}[X], T_2^{\alpha}[X]) = \beta$. Since r satisfies $X \sim\to\sim\to_\alpha Y$, there must be a tuple T in r such that: (V=R-XY)

    $SP^{\alpha}(T^{\alpha}[X], T_1^{\alpha}[X]) \geq \beta$
    $SP^{\alpha}(T^{\alpha}[X], T_2^{\alpha}[X]) \geq \beta$
    $SP^{\alpha}(T^{\alpha}[Y], T_1^{\alpha}[Y]) \geq \beta$
    $SP^{\alpha}(T^{\alpha}[V], T_2^{\alpha}[V]) \geq \beta$

    Since $Z \cap Y = \phi$, $Z \subseteq XV$, hence $SP^{\alpha}(T^{\alpha}[Z], T_2^{\alpha}[Z]) \geq \beta$. The FFD $Z \sim\to_\alpha W$ means that $SP^{\alpha}(T^{\alpha}[W], T_2^{\alpha}[W]) \geq \beta$. However, since $W \subseteq Y$, therefore we can conclude $SP^{\alpha}(T^{\alpha}[W], T_1^{\alpha}[W]) \geq \beta$, and as a result r will satisfy $X \sim\to_\alpha W$.

**Theorem 3**: The inference rules based on the concept of extension of semantic proximity are complete.

**Proof**: The article [6] proved the completeness of inference rules for FFD at degree α. Now we want to prove the completeness of inference rules for FMVD at degree α.

Lets assume F and G are given sets of FFDs and FMVDs on a relation r on R. The closure of F U G, denoted by $(F,G)^+$, is the set of all FFDs and FMVDs that can derive inference rules from F U G. Let X be a subset of R. There are several sets Y such that the FMVD $X \sim\rightarrow\sim\rightarrow_\alpha Y$ is in $(F,G)^+$. For example, $X \sim\rightarrow\sim\rightarrow_\alpha X$, and $X \sim\rightarrow\sim\rightarrow_\alpha R-X$ are always in $(F,G)^+$. Following classical relational database literature we use the notation $X \sim\rightarrow\sim\rightarrow_\alpha Y_1 | Y_2 |...| Y_n$. To denote the collection of FMVDs $X X \sim\rightarrow\sim\rightarrow_\alpha Y_1, X \sim\rightarrow\sim\rightarrow_\alpha Y_2,..., X \sim\rightarrow\sim\rightarrow_\alpha Y_n$, and none of these $Y_1,…,Y_n$ are empty. The Additivity and Projectivity rule of FMVD lets us partition R into sets of attributes $Y_1,…,Y_n$, so that, the given dependencies can be derived by Additivity rule.

**Theorem 4:** A FFD can be a case of FMVD.

**Proof**: According to Replication rule, if r satisfies the FFD, then r will satisfy the FMVD. We can also conclude that a FFD is a case of FMVD.

**Theorem 5:** A classical MVD satisfies the definition of FMVD at degree α.

**Proof**: Since the extension of semantic proximity supports Crisp values, then the definition of FMVD at degree α supports them. Therefore, a classical MVD satisfies the definition of FMVD at degree α.

We show a brief comparison between FMVD at degree α, FMVD based on definition 1 and FMVD based on definition 5 in Table 2:

Table 2: A brief comparison between different FMVD definitions

|  | **Crisp and NULL Values** | **Fuzzy Values** | **Fuzzy values have some out of the range data** |
|---|---|---|---|
| **FMVD based on definition 1** | Does not work | Works | Does not work |
| **FMVD based on definition 5** | Works | Works | Does not work |
| **FMVD at degree α** | Works | Works | Works |

## 6. CONCLUSIONS

In this article, we developed a new approach for solving the problems with Crisp and Null values that exist in the concept of semantic proximity. We further defined FMVD based on an extension of semantic proximity at degree α for fuzzy relational databases that have Crisp, NULL and fuzzy values and reduce redundancy and anomaly in the databases. The problem with fuzzy values in the form of interval numbers, that might still have some out of the range data, is also solved by defining FMVD. Furthermore, inference rules are presented and proved for FMVD and FMVD-FFD as well as shown to be complete. Finally, based on the extension of semantic proximity concept, we show FFD at degree α and MVD are special cases of FMVD.

International Journal of Database Management Systems ( IJDMS ), Vol.3, No.3, August 2011## REFERENCES

## AUTHORS

**Arezoo Rajaei** was born in Shahreza, Isfahan, Iran. She received a B.S. degree in Computer Engineering in 2007, from Ferdowsi University of Mashhad, Iran. She is currently an M.S. degree candidate in Computer Engineering from SheikhBahaee University,Isfahan, Iran. Her research interests include fuzzy database and genetic algorithm.

**Ahmad Baraani Dastjerdi** received his B.S. degree in Statistics and Computing from Ferdowsi University of Mashhad in 1977, an M.S. degree in Computer Science from George Washington University, USA in 1979, and a Ph.D. in Computer Science from University of Wollongong, Australia in 1996. His expertise and interests   include databases, Data security, Information Systems, and e-Society. He is currently an assistant professor of computer engineering at Isfahan University, Iran.

**Nasser Ghasem Aghaee** received his B.S. degree in Mathematics from Isfahan University in 1972, an M.S. degree in Information and Computer Science from Georgia Institute of Technology, USA, in 1977, and a Ph.D. in Computer Science from University of Bradford, U.K., in 1996. His expertise and interests include Computer Simulation, Advance Artificial Intelligence, and User Modeling. He is currently professor of computer engineering at SheikhBahaee University & Isfahan University, Iran